# Compressive Diffraction Tomography for Weakly Scattering

Lianlin Li,    Wenji Zhang,    Fang Li

Institute of Electronics, Chinese Academy of Sciences, Beijing, China

*Lianlinli1980@gmail.com*

*ABSTRACT-*

An appealing requirement from the well-known diffraction tomography (DT) exists for success reconstruction from few-view and limited-angle data. Inspired by the well-known compressive sensing (CS), the *accurate super-resolution* reconstruction from highly sparse data for the weakly scatters has been investigated in this paper. To realize the "compressive" data measurement, in particular, to obtain the super-resolution reconstruction with highly sparse data, the "compressive" system which is realized by surrounding the probed obstacles by the random media has been proposed and empirically studied. Several interesting conclusions have been drawn: (*a*) if the desired resolution is within the range from $0.2\lambda$ to $0.4\lambda$, the *K*-sparse *N*-unknowns imaging can be obtained exactly by $O\left(K\log\left(N/K\right)\right)$ measurements, which is comparable to the required number of measurement by the Gaussian random matrix in the literatures of compressive sensing. (*b*) With incorporating the random media which is used to enforce the multi-path effect of wave propagation, the resulting measurement matrix is incoherence with wavelet matrix, in other words, when the probed obstacles are sparse with the framework of wavelet, the required number of measurements for successful reconstruction is similar as above. (*c*) If the expected resolution is lower than $0.4\lambda$, the required number of measurements of proposed "compressive" system is almost identical to the case of free space. (*d*) There is also a requirement to make the tradeoff between the imaging resolutions and the number of measurements. In addition, by the introduction of complex Gaussian variable the kind of fast sparse Bayesian algorithm has been slightly modified to deal with the complex-valued optimization with sparse constraints.

*INDEX TERM*-

Inverse scattering problem, the diffractive tomography(DT), the Born/Rytov approximation, the super-resolution imaging, the sparse Bayesian optimization, compressive sensing (CS),

## I.    INTRODUCTION

In the well-known diffraction tomography (DT), the electrical and geometrical properties of an object can be exactly reconstructed from the object's diffracted field.

This modality has been extensively investigated for its potential applications to optical imaging, medical imaging, subsurface imaging, geophysical imaging, weather prediction, ionosphere tomography, and so on. Under the assumption of weak scattering one can use the Born or Rytov approximations to derive the so-called Fourier diffraction projection theorem (FDPT), which relates the 2D Fourier transform of the object to the measured data. The input data to the FDPT algorithm and its all kinds of variants should satisfy the so-called Nyquist-Shannon theorem, the full-view and full-angle measurements tightly packed over the sphere with radius $k$ ($k$ denotes the working wavenumber of background media), in order to obtain the desired reconstruction. On the other hand, due to the use of Fourier transform the theoretical limit of resolution by FDPT is $\lambda/2$ ($\lambda$ is the working wavelength), which means that the FDPT algorithm excludes the possibility of resolving, under realistic measurement noise levels, features of a wave scattering object which are less than $\lambda/2$ apart. Consequently, two appealing problems for the DT are that how to reduce the number of measurements (henceforce, to decrease the data-collecting time) and how to improve the imaging resolution.

There has been recent effort to develop the super-resolution reconstruction algorithms from few-view limited-angle data (see [10-14] and references in there). In [12], by taking the rich information of decaying wave in the region of near field into account, the super-resolution imaging based on the Born-based inverse scattering algorithm with near-field data has been investigated, by which the imaging with sub-wavelength resolution can be obtained, even in the presence of large noise levels. The further insight into the mechanism which leads to sub-wavelength resolution imaging from far-field measurement by means of multiple scattering has been carried out experimentally and theoretically, respectively, for example, [13] and [14] and others. In [10], the sparse prior information on the gradient-magnitude images (GMIs) has been exploited to obtain the full Fourier reconstruction from few-view and limited-angle Fourier data compatible with the assumptions of the FDPT, in particular, the proposed algorithm iteratively minimizes the total variation (TV) of the estimated image subject to the constraint that the reconstructed image matches the measured object data.

For the purpose of the sub-wavelength DT reconstruction from few-view limited-angle data, the linear complex-valued equations to be solved are highly underdetermined. Obviously, there are infinitely many solutions. To attack it, the additional, *reasonable* and *universal* prior information on solution should be enforced to guarantee a unique solution; furthermore, the corresponding tractable fast algorithms should be developed. The well-known compressive sensing theory developed recently by Candes, Romberg and Tao et al, of course, and the variants of it claim that when the solution is enough sparse and the matrix forming the linear equations possesses of some properties, for example, the so-called restricted isometry property, the Null-space property, and so on, the solution can be exactly, stably and uniquely determined, even by the greedy algorithm. Three key issues of compressive sensing are (1)sparse transformation matrix by which the information of unknown signal can be captured by much smaller coefficients than the dimension of original signal, (2)the measurement matrix incoherence with transformation matrix and (3) the optimization algorithm with the sparse constraints. To achieve them, the key is to construct a *universal* measurement matrix which should at least satisfy (a) it should succeed using a minimal number of samples, (b) it should be robust when samples are contaminated with noise, (c) it should provide optimal error guarantees for every target signal. Candes and Tao shown that by the Gaussian random matrices the measurements with the order of $O(K\log(N/K))$ is sufficient to exactly reconstruct a *K*-sparse *N*-dimensional signal by solving a convex optimization problem. Even more, Tropp, Gilbert and coworkers shown that by running the greedy algorithm over the problem, the measurements with the order of $O(K\log(N))$ is sufficient to finish the same task as above. Of course, there are many other excellent results about this topic.

To realize the "compressive" data measurement or construct the good measurement matrix, in particular, to obtain the super-resolution reconstruction with highly sparse data, the "compressive" system which is realized by surrounding the probed obstacles by the random media has been proposed and empirically investigated. Taking this into account, the term of compressive diffraction tomography is called in this paper. The remainder of this paper is arrangement as following. In Section 2, the compressive diffractive tomography has been presented and investigated, by which the super-resolution imaging can be reconstructed from few-view data and limited- view scan data. In Section 3, the complex-valued fast sparse Bayesian algorithm has been provided. Finally some numerical simulations are shown.

## II. METHODS

We consider two-dimensional scattering to simplifying notation, although the basic construct extends naturally to three-dimensional full-vectorial problems. As discussion in section I, one of the key issues of compressive diffraction tomography or compressive sensing is to construct the suitable *universal* measurement matrix incoherence with arbitrary sparse transformation such as DCT, wavelet, and so on. Moreover, by using this measurement, the required number of measurements is minimal. To do this, the complex electromagnetic environment where the probed obstacles are located has been constructed by adding the random media to enforce the significant multi-path effect, such as that considered in previous time-reversal studies (see Fig. 1). Thereby, the objective is to exploit the multi-path effect introduced by such an environment to obtain the super-resolution reconstruction from the relatively small number of measurements performed in the presence of the heterogeneities, using ideas from compressive sensing. From the results below, one can find that the resulting matrix for the compressive diffractive tomography can be comparable to the Gaussian random matrix appeared in the literatures of compressive sensing.

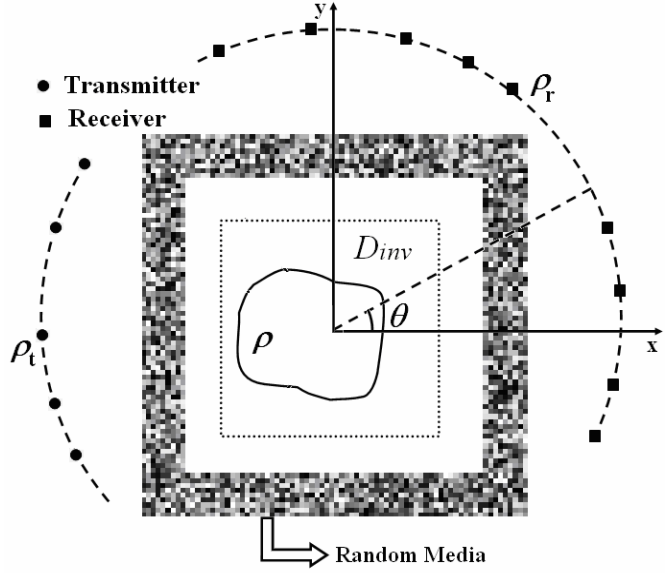

FIG. 1 The sketch map of compressive diffraction tomography

Under the Born approximation for weakly scatters, the desired signal $E_s(\bar{r}_R, \bar{r}_S; k)$, representing the scattered fields observed at $\bar{r}_R$ ($R = 1, 2, \cdots, N_R$, $N_R$ is the number of receivers) due to incident filed from the source at $\bar{r}_S$ ($S = 1, 2, \cdots, N_S$, $N_S$ is the number of transimitters), can be expressed as

$$E_s(\bar{r}_R, \bar{r}_S; k) = k^2 \int_{Dinv} G(\bar{r}_R, \bar{r}'; k) E_{in}(\bar{r}_S, \bar{r}'; k) \chi(\bar{r}') d\bar{r}' \qquad (2.1)$$

where $\chi(\bar{r}') = 1 - \dfrac{k(\bar{r}')}{k}$, $k$ is wavenumber in free space, $k(\bar{r}')$ denotes the wavenumber of the unknown object, $E_{in}(\bar{r}_S, \bar{r}'; k)$ denotes the incident wave due to the source at $\bar{r}_S$, $G(\bar{r}_R, \bar{r}'; k)$ is the Green function for the random background media. It is noted that $G(\bar{r}_R, \bar{r}'; k)$ can be obtained off-line by the existing methods such as method of moment, the finite element method, and so on, and stored.

Using the similar notations employed in compressive sensing, we rewrite (2.1) in the form of the matrix-vector as

$$\bar{y} = \bar{\bar{A}}\bar{x} + \bar{n}, \qquad \bar{n}, \bar{y} \in \mathbb{C}^{M \times 1}, \ \bar{\bar{A}} \in \mathbb{C}^{M \times N}, \ \bar{x} \in \mathbb{C}^{N \times 1} \qquad (2.2)$$

where the entries of $\bar{y}$ is from $E_s(\bar{r}_R, \bar{r}_S; k)$, $\bar{n}$ denotes the error due to measurement noise and discretization transformation from (2.1) to (2.2), the entries of $\bar{\bar{A}}$ are the form of $\Delta k^2 G(\bar{r}_R, \bar{r}_j; k) E_{in}(\bar{r}_S, \bar{r}_j; k)$ ($j = 1, 2, \cdots, N$, $N$ is the total number of the unknowns to be reconstructed), the entries of $\bar{x}$ is $\chi(\bar{r}'_j)$. Further, if $\bar{x}$ can be sparse in some orthogonal basis $\bar{\bar{\Psi}}$ such as the wavelet transform, the discrete cosine transform (DCT), and so on, in particular, $\bar{x} = \bar{\bar{\Psi}} \bar{\theta}$, the equation (2.2) can be rewritten as

$$\bar{y} = \bar{\bar{A}} \bar{\bar{\Psi}} \bar{\theta} + \bar{n}, \qquad \bar{n}, \bar{y} \in \mathbb{C}^{M \times 1}, \ \bar{\bar{A}} \in \mathbb{C}^{M \times N}, \ \bar{\theta} \in \mathbb{C}^{N \times 1} \qquad (2.3)$$

For the purpose of the super-resolution reconstruction from the few-view limited-angle measurements, the formed complex-valued matrix $\bar{\bar{A}}$ is highly underdetermined, i.e., $M \ll N$. Obviously, the highly undermined system of linear equations (2.2) or (2.3) will generates infinitely many solutions. If one desires to narrow the choice to one well-defined solution, additional criteria should be needed to guarantee a unique solution. To attack this, define the general optimization problem

$$\hat{x} = \min_{\bar{x}} J(\bar{x}), \quad \text{s.t.} \quad \|\bar{y} - \bar{\bar{A}}\bar{x}\|_2 \leq \|\bar{n}\|_2 \qquad (2.4)$$

where $J(\bar{x})$ is introduced to evaluate the desirability of a would-be solution $\bar{x}$. Usually, the squared $\ell_2$ norm $J(\bar{x}) = \|\bar{x}\|_2$, a measure of energy, is employed; further, the so-called Tikhnov regularization is carried out to obtain a unique solution. It is well known that the measure of energy is not optimal, even worst in some cases. When the minimal squared $\ell_2$ norm is carried out, the values of non-measured scattered field are implicitly enforced to zero. Obviously, it is unreasonable in most of cases. So we should find more suitable constraints on $\bar{x}$ than squared $\ell_2$ norm in order to obtain or improve the reconstruction. Fortunately, all most of obstacles to be probed are with finite support, in other words, sparse, or compressible in some framework such as DCT, wavelet, and so on. Naturally, to seek the sparest solution to (2.2) or (2.3) become an appealing problem. In recent years, finding the sparest solution to highly undetermined linear equations has been explored in depth and numerous surprising results have been drawn, especially, since the so-called compressive sensing was developed, whether theory or algorithm.

    Below we will show numerically that the resulting matrix possesses of the good properties of the Gaussian random matrix appeared in the literatures of compressive sensing. As mentioned above, one of key issues of compressive sensing is to construct a suitable measurement matrix. Two important properties of good measurement

matrix are that (1) it is *universal*, in the notation of compressive sensing, it is incoherence with any arbitrary orthogonal basis. In particular, the unknown $\bar{x}$ which is sparse in the arbitrary orthogonal basis can be reconstructed exactly. (2) The required number of measurements is minimal. Due to the lack of theoretical analysis on the Green' functions for random media, the empirical analysis based on numerical simulations are carried out in our work, by which a general conclusion has been drawn

. For our numerical experiments, we assume the aforementioned excitation is a line current, but the formulation generalizes to arbitrary two/three-dimensional sources. The transmitters are located uniquely distributed from 0° to 120° with separation 40°, for each transmitter, the 16 receivers are uniquely located the range from 0° to 75°. The size of investigation is 128 by 256 with different grid size for the purpose of studying the different reconstruction resolution. Consequently, the size of considering matrix $\bar{\bar{A}}$ shown in equation (2.2) is 128 by 256. Moreover, the simulated data computed by the moment method is contaminated by 30dB SNR Gaussian noise. As done in compressive sensing, for each of 40 trials we randomly choose the support of $\bar{x}$ for each value of $K$ ($K$ denotes the number of nonzeros of $\bar{x}$), in particular, choose the sign of the components from a Gaussian distribution of mean 0 and standard derivation 1(the MATLAB code is: *tmp1=randperm(N); tmp2=tmp1(1:K); x(tmp2)=sign(randn(K,1))*). Results are shown in Fig. 2 for different reconstruction resolution where the sparse transformation matrix is unit matrix, $(a) 0.1\lambda$, $(b) 0.2\lambda$, $(c) 0.3\lambda$, $(d) 0.4\lambda$, $(e) 0.5\lambda$ and $(f) 0.6\lambda$. To compare the properties of matrix $\bar{\bar{A}}$ of compressive diffraction tomography with one of Gaussian random matrix appeared in the literature of compressive sensing, we provides the plots of rate of success as the function of $K$ in Fig. 2(*g*). Moreover, in order to check the role of random media to enhance the "compressive" data, the plots of recovery frequency as a function of $K$ for free-space background has also been provided, represented by the blue line in stead of red line. The plot of knee points of Figs.2 and reconstruction resolution is shown in Fig.3. For all figures, the *x*-axis denotes the $K$ while *y*-axis denotes the rate of success. From the Figs. 2 and Fig.3, one can readily draw the following conclusions:

(1) Though the measurement matrix for the free space also guarantee the success reconstruction with sparse data, the random media does have the strong ability of compressing data, especially, for the desired resolution within $0.2\lambda$ and $0.5\lambda$.

(2) For the considering sparse signal, in particular, the value of entries of $\bar{x}$ is $(0, \pm 1)$, the required number of measurements for the *exact* reconstruction will be $O\left(2K \log\left(N/K\right)\right)$, comparable to the case of Gaussian random matrix in the literature of compressive sensing, especially, for the case of the $0.3\lambda$ reconstruction resolution.

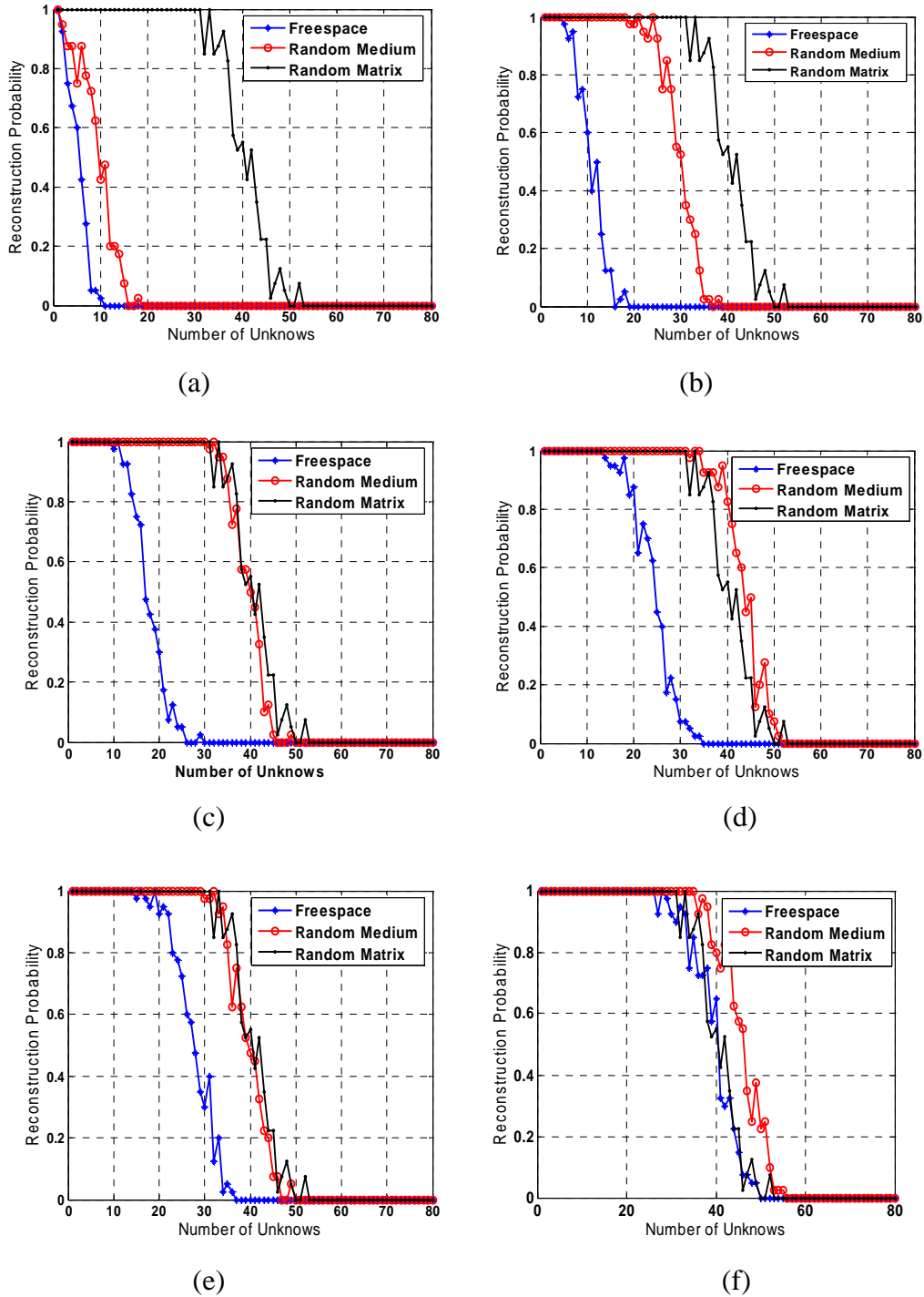

FIG.2 Plots of recovery frequency as a function of *K* for different resolution, where red-line is for compressive diffraction tomography, the blue-line is free-space diffraction tomography. (a) $0.1\lambda$, (b) $0.2\lambda$, (c) $0.3\lambda$, (d) $0.4\lambda$, (e) $0.5\lambda$ and (f) $0.6\lambda$. For comparison, the plots of recovery frequency as a function of *K* for Gaussian matrix has been provided.

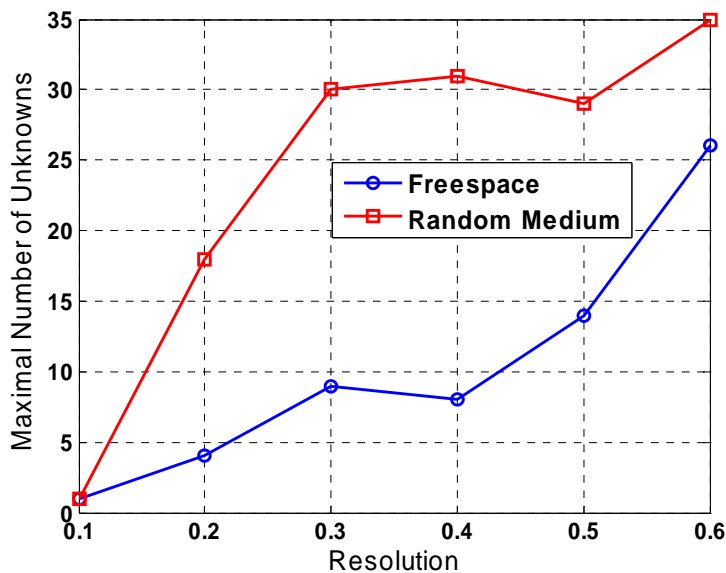

Fig.3 The curve of relation between resolution and sparsity *K* for compressive diffraction tomography (the red-line) and the free-space diffraction tomography (the blue line)

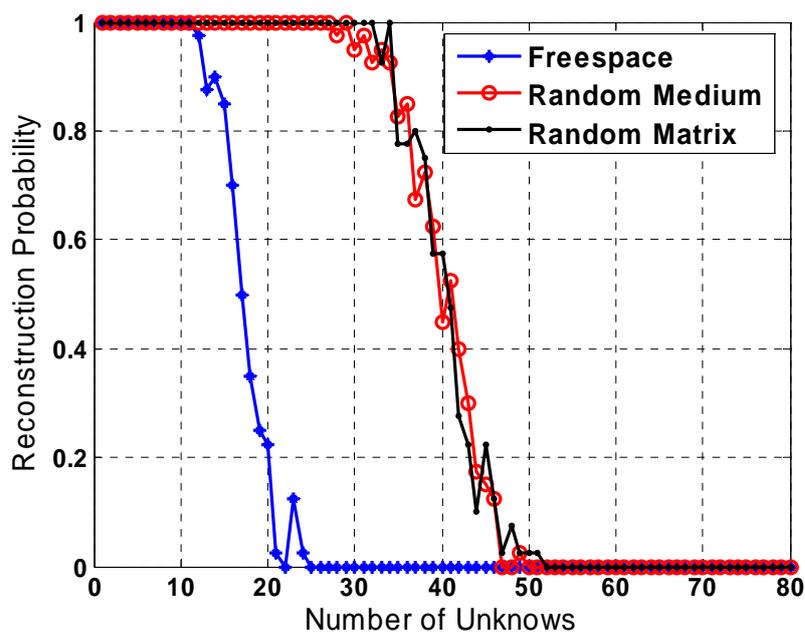

FIG. 4 Plots of recovery frequency as a function of *K* with $0.3\lambda$ reconstruction resolution for Haar-wavelet sparse transformation, where red-line is for compressive diffraction tomography, the blue-line is free-space diffraction tomography.

To check the universality of the matrix of compressive diffraction tomography for different sparse transformation, Taking reconstruction with $0.3\lambda$ resolution as example, Fig.4 shows the plots of rate of success as the function of *K* for Haar-wavelet. In addition, the similar result for free-space diffraction tomography has been provided. Fig.4 shows that the matrix of compressive diffraction tomography is universal, especially incoherence with Haar-wavelet; much better than one of free-space diffraction tomography.

### III. THE COMPLEX-VALUED FAST BAYESIAN ALGORITHM WITH THE SPARSE CONSTRAINTS

There have been limited investigation on the compressive sensing inversion based on complex-valued data, and here we extend the Bayesian compressive sensing algorithm discussed in [6-8] to the complex-valued case by introducing the signal/measurement model of the complex Gaussian distribution. Consider the highly underdetermined complex-valued linear equations

$$\bar{y} = \bar{\bar{\Phi}}\bar{w} + \bar{n}, \qquad (3.1)$$

with $\bar{\bar{\Phi}} \in \mathbb{C}^{M \times N}$, $\bar{w} \in \mathbb{C}^{N \times 1}$ and $\bar{y}, \bar{n} \in \mathbb{C}^{M \times 1}$.

To carry out the Bayes analysis, the two models should be assumed, in particular,

(1) THE DATA MODEL.

The observation noise $\bar{n}$ is independent and complex Gaussian distribution $\mathcal{N}_c$ with zero mean and variance equal to $\beta$, that is,

$$\Pr(\bar{y} | \bar{w}) = \mathcal{N}_c\left(\bar{y} | \bar{\bar{\Phi}}\bar{w}, \bar{\bar{\Sigma}}_n\right)$$
$$= \left(\frac{1}{2\pi}\right)^{M/2} \left|\bar{\bar{\Sigma}}_n\right|^{-0.5} \exp\left(-\frac{1}{2} \left\|\bar{y} - \bar{\bar{\Phi}}\bar{w}\right\|^2_{\Sigma_n^{-1}}\right) \qquad (3.2)$$

where $\bar{\bar{\Sigma}}_n = \beta \bar{\bar{I}}$.

(2) THE SPARSE-SIGNAL MODEL.

The intuition constraint on $\bar{w}$ is the so-called Laplace prior, i.e. $\Pr(w_i | \lambda_i) = \frac{\lambda_i}{2} \exp\left(-\frac{\lambda_i}{2} |w_i|\right)$, where $\lambda_i \geq 0$. However, this Laplace prior does not allow for a tractable Bayesian analysis, since it is not conjugate to the conditional distribution in (3.2). To alleviate this, a common strategy is the hierarchical prior. In following, the proposed procedure [] is borrowed, where the only difference is that the considered variables are complex-valued.

The first stage of a hierarchical model is that the prior on $\bar{w}$ is the complex Gaussian distribution with zero mean and variance $\gamma_i$, in particular,

$$\Pr(\bar{w} | \bar{\gamma}) = \prod_{i=1}^{N} \mathcal{N}_c(w_i | 0, \gamma_i) \qquad (3.3)$$

In the second stage of the hierarchy, the Gamma distribution with $(1, \lambda/2)$ is utilized independently on each $\gamma_i$, that is

$$\Pr(\gamma_i | \lambda) = Gamma(\gamma_i | 1, \lambda/2) = \frac{\lambda}{2} \exp\left(-\frac{\lambda}{2}\gamma_i\right) \tag{3.4}$$

It is noted that from (3.3) and (3.4) the resulting distribution $\Pr(\bar{w} | \lambda)$ is a Laplace distribution, in particular,

$$\Pr(\bar{w} | \lambda) = \int \Pr(\bar{w} | \bar{\gamma}) \Pr(\bar{\gamma} | \lambda) d\bar{\gamma} = \frac{\lambda^{N/2}}{2^N} \exp\left(-\sqrt{\lambda} \sum_i |w_i|\right) \tag{3.5}$$

The third stage of the hierarchy is the Gamma hyperprior with $(v/2, v/2)$ on $\lambda$, in particular,

$$\Pr(\lambda | v) = Gamma(\lambda | v/2, v/2) \tag{3.6}$$

Now, equations (3.3) (3.4) and (3.5) finish the proposed the three-stage hierarchical model of sparse signal. The incoming key issue is to find the $\bar{w}$ under the condition of $\bar{y}, \beta, \bar{\gamma}$ and $\lambda$. According to Bayes's rule, one has

$$\Pr(\bar{w} | \bar{y}, \beta, \bar{\gamma}, \lambda) = \frac{\Pr(\bar{w}, \bar{y}, \beta, \bar{\gamma}, \lambda)}{\Pr(\bar{y}, \beta, \bar{\gamma})} \sim \Pr(\bar{y} | \beta, \bar{w}) \Pr(\bar{w} | \bar{\gamma}, \lambda)$$

After some simple operations, one has from above equations

$$\Pr(\bar{w} | \bar{y}, \beta, \bar{\gamma}, \lambda) \sim \mathcal{N}_c(\bar{y} | 0, \bar{\bar{C}}) \mathcal{N}_c(\bar{w} | \bar{\mu}, \bar{\bar{\Sigma}}) \tag{3.7}$$

where

$$\bar{\mu} = \beta \bar{\bar{\Sigma}} \bar{\bar{A}}^H \bar{y}, \qquad \bar{\bar{\Sigma}} = \left(\beta \bar{\bar{A}}^H \bar{\bar{A}} + \bar{\bar{\Lambda}}^{-1}\right)^{-1}$$

$$\bar{\bar{C}} = \beta^{-1} \bar{\bar{I}} + \bar{\bar{A}} \bar{\bar{\Lambda}} \bar{\bar{A}}^H, \qquad \bar{\bar{\Lambda}} = diag(\gamma_i)$$

From (3.7) one easily has the optimal solution to (3.1) is $\hat{x} = \bar{\mu}$. However, it is noted that $\bar{\mu}$ is highly dependent on the choice of $\bar{\bar{\Lambda}}$. Then our next task is to find the suitable $\bar{\bar{\Lambda}}$ within the framework of Bayes. By using (3.7) and

$$\Pr(\bar{\gamma}, \beta, \lambda | \bar{y}) = \frac{\Pr(\bar{\gamma}, \beta, \lambda, \bar{y})}{\Pr(\bar{y})} \propto \Pr(\bar{y}, \bar{\gamma}, \beta, \lambda),$$

one has

$$\Pr(\bar{y},\bar{\gamma},\beta,\lambda) = \int \Pr(\bar{y}|\bar{w},\beta)\Pr(\bar{w}|\bar{\gamma})\Pr(\bar{\gamma}|\lambda)\Pr(\lambda)\Pr(\beta)d\bar{w}$$

$$= \left(\frac{1}{2\pi}\right)^{N/2} |\bar{\bar{C}}|^{-0.5} \exp\left(-\frac{1}{2}\bar{y}^H \bar{\bar{C}}^{-1}\bar{y}\right) \Pr(\bar{\gamma}|\lambda)\Pr(\lambda)\Pr(\beta) \quad (3.8)$$

As done by [6-8], one has

$$\mathcal{L}(\bar{\gamma}) = \log \Pr(\bar{y},\bar{\gamma},\beta,\lambda)$$
$$= \mathcal{L}(\bar{\gamma}_{-i}) + l(\bar{\gamma}_i)$$

with $l(\bar{\gamma}_i) = \frac{1}{2}\left[\log\left(\frac{1}{1+\gamma_i s_i}\right) + \frac{\gamma_i |q_i|^2}{1+\gamma_i s_i} - \lambda \gamma_i\right]$,

$$s_i = \bar{\phi}_i^H \bar{\bar{C}}_{-i}^{-1} \bar{\phi}_i$$

$$q_i = \bar{\phi}_i^H \bar{\bar{C}}_{-i}^{-1} \bar{y}$$

It is noted that $s_i, i=1,2,\cdots,N$ are real-valued variables. Now, one can obtain the value of hyper-parameters by solving $\frac{d}{d\bar{\gamma}_i}\mathcal{L}(\bar{\gamma})=0$, in particular,

$$\gamma_i = \begin{cases} \dfrac{-s_i(s_i+2\lambda) + s_i\sqrt{(s_i+2\lambda)^2 - 4\lambda(s_i-|q_i|^2+\lambda)}}{2\lambda s_i^2} & |q_i|^2 - s_i > \lambda \\ 0 & else \end{cases}$$

As for the detailed discussion, we would like to refer the readers to [8].

## IV. NUMERICAL SIMULATIONS

In this section we would to perform two sets of numerical simulations to demonstrate and validate above theory and algorithm, where the investigation domain with size *3m* by *3m* has been divided into the subgrids with size $0.1\lambda$ by $0.1\lambda$ (see Fig.1), other computation condition is the same as one given in section II.

*Numerical Example 1: Reconstruction with unit-matrix transformation*

For this example, the probed objects are assumed to be as sparse itself (see Fig. 5(a)). The reconstructed results are shown in Figs. 5, where (b) is the result of free-space diffraction tomography by $\ell_2$-norm constraint, (c)is the result of free-space diffraction tomography by complex-valued sparse Bayesian algorithm, (d) is the result of compressive diffraction tomography by $\ell_2$-norm constraint, (e) is the result of compressive diffraction tomography by complex-valued sparse Bayesian algorithm.

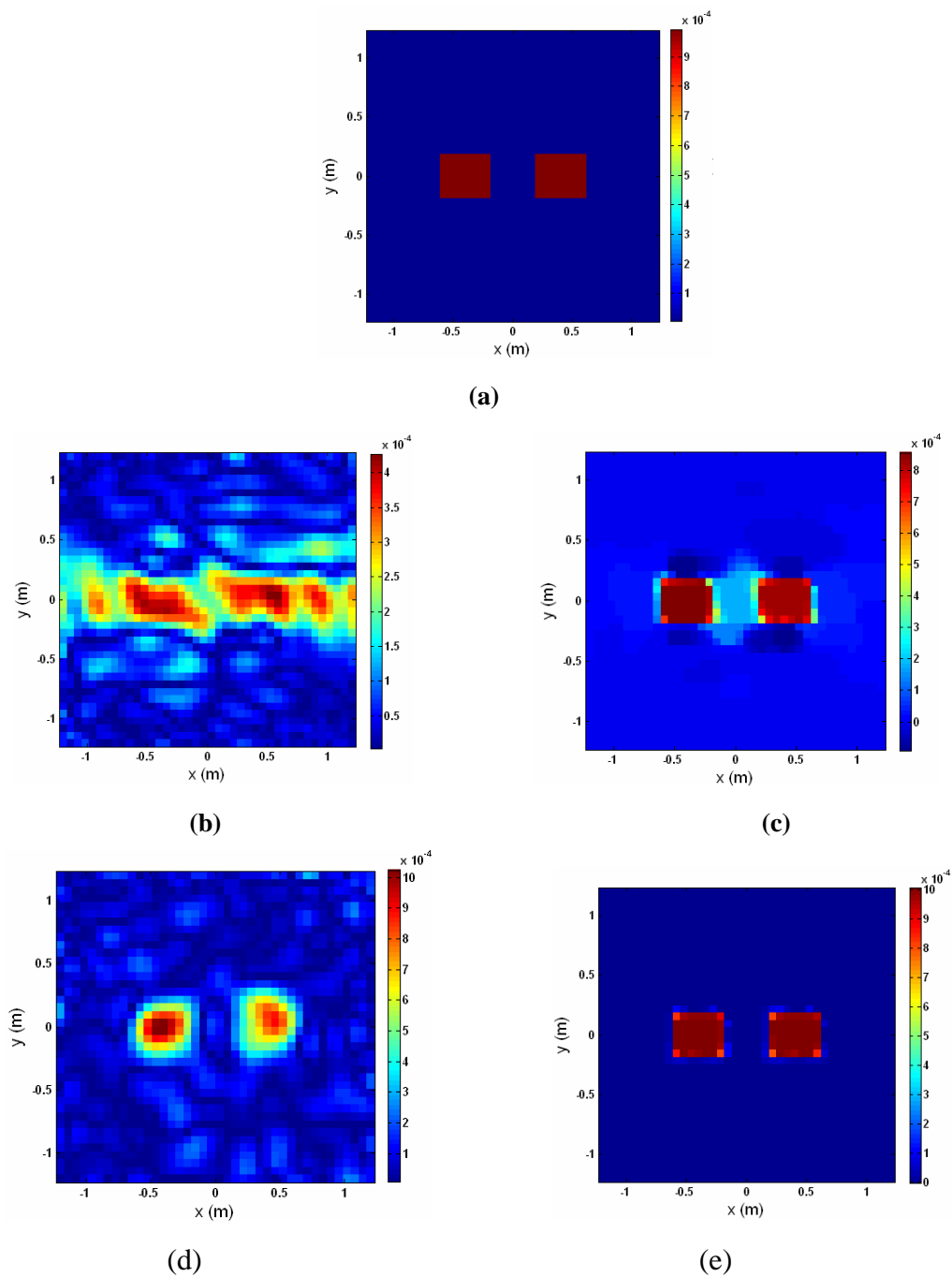

FI.G 5 Diffraction tomography reconstruction with unit-matrix transformation from sparse data (b) free-space background with $\ell$-2 constraints optimization, (c) free-space background with sparse Bayesian optimization, (d) random-media background with $\ell$-2 constraints optimization and (*e*) random-media background with sparse Bayesian optimization.

*Numerical Example 2: Reconstruction with Haar-wavelet transformation*

For this example, the probed obstacles are sparse in the framework of Haar-wavelet transformation (see Fig.6(a)). The reconstructed results are shown in Figs. 5, where (b) is the result of free-space diffraction tomography by $\ell$-2 constraint, (c)is the result of free-space diffraction tomography by complex-valued sparse Bayesian algorithm, (d) is the result of compressive diffraction tomography by $\ell$-2-norm constraint, (e) is the result of compressive diffraction tomography by complex-valued sparse Bayesian algorithm.

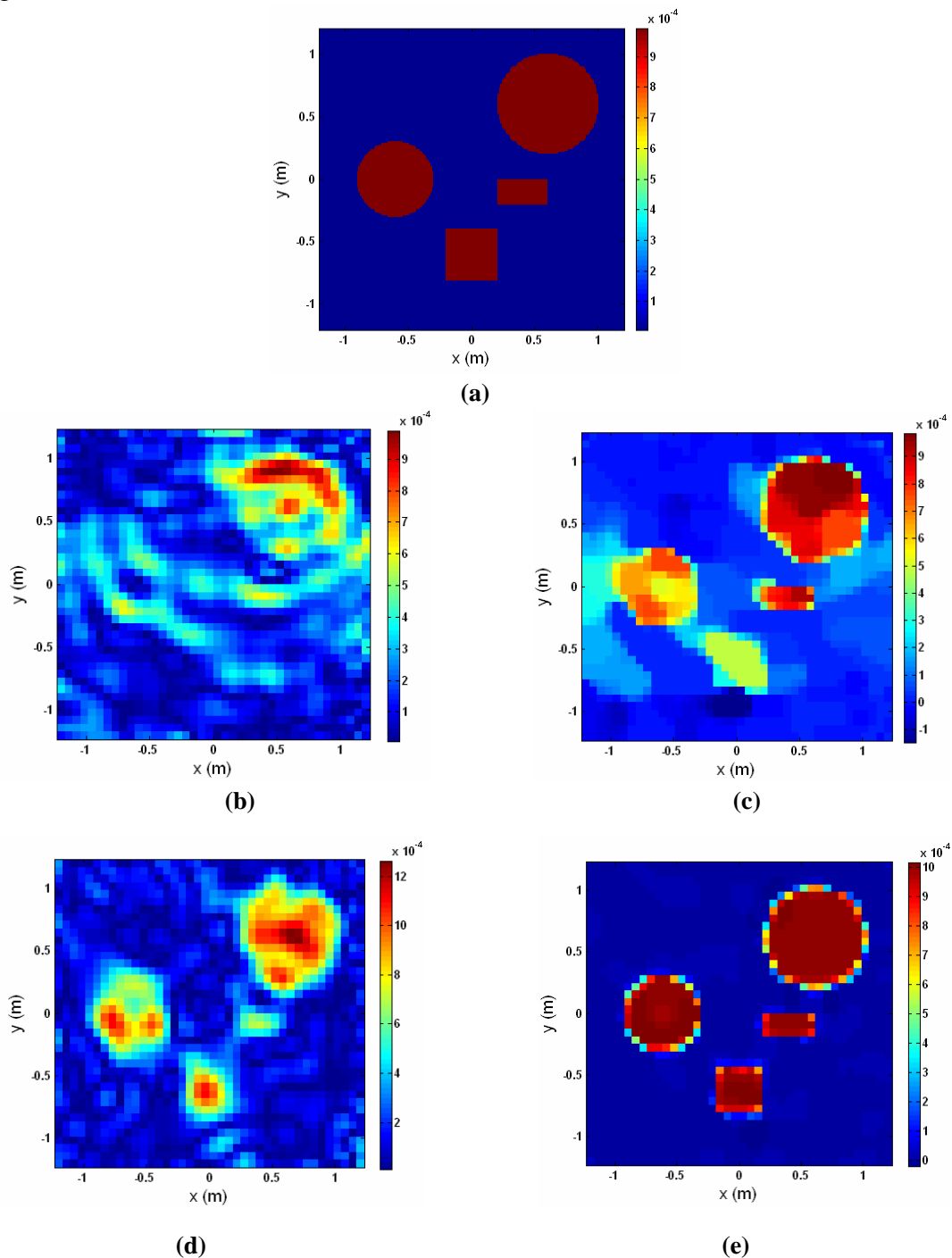

FI.G 6 Diffraction tomography reconstruction with Haar-wavelet transformation from sparse data (b) free-space background with $\ell$-2 constraints optimization, (c) free-space background with sparse Bayesian optimization, (d) random-media background with $\ell$-2 constraints optimization and (e) random-media background with sparse Bayesian optimization.

From Figs.5 and Figs.6, one can find that (1) The reconstruction for the free-space diffraction tomography can not also yield visually reasonable reconstructions; while the reconstruction by compressive diffraction tomography can do this. (b) Although the reconstruction for the $l_2$-constraint and random media background media can also yield visually reasonable reconstructions, its results are much poorer than the reconstruction with sparse Bayesian algorithm. (c) Among all results, the reconstruction by random media background and sparse Bayesian algorithm is the best, which proves that one can obtain the super-resolution imaging from highly sparse (few-view and limited-angle) data. We have also performed tests using a series of different scan configurations, trading off between decreasing the number of views and increasing the number of samples per view, in an effort to demonstrate accurate image reconstruction from extremely sparse data. In summary, by the random-media background and sparse-enhance optimization, one can obtain exactly super-resolution reconstruction from the highly sparse data. As notation by compressive sensing, this diffraction tomography is called compressive diffraction tomography (CDT).

## IV. CONCLUSIONS

In this paper, a novel super-resolution diffraction tomography from highly sparse data, called compressive diffraction tomography in this paper, has been proposed and analyzed. A interesting and important conclusion has been empirically drawn that by using the compressive diffraction tomography one can realize the exact super-resolution reconstruction of compressible obstacles from $O\left(K \log\left(N/K\right)\right)$ measurements, whether objects with finite support or one compressible in some orthogonal basis. As for the assumption of compressible object, this is a reasonable assumption for many practical applications such as medical imaging, nondestructive test, subsurface imaging, and so on. There is also a requirement to make the tradeoff between the imaging resolutions and the number of measurements. Of course, how to explain strictly this should be made, which will be the direction of future research. Moreover, by the introduction of complex Gaussian variable the kind of fast sparse Bayesian algorithm has been slightly modified to deal with the complex-valued optimization with sparse constraints.

ACKNOWLEGE:
This work has been supported by the National Natural Science Foundation of China under Grants 60701010 and 40774093.